\documentclass[prl,superscriptaddress,twocolumn]{revtex4}
\usepackage{amssymb}
\usepackage{amsmath}
\usepackage{amsfonts}
\usepackage{fixmath}
\usepackage{graphicx}
\usepackage{mathrsfs}
\newcommand{\SU}[1]{\mathbb{SU}(#1)}
\newcommand{\ket}[1]{|#1\rangle}
\newcommand{\bra}[1]{\langle#1|}
\newcommand{\Tr}{\textrm{Tr}}
\newcommand{\ceq}{\mathrel{\mathop:}=}
\renewcommand{\vec}[1]{\boldsymbol{#1}}
\def\map#1{{\mathscr{#1}}}
\begin{document}
\title{Erasable and unerasable correlations}

\author{G. M. D'Ariano} 

\affiliation{QUIT Group, Dipartimento di Fisica ``A. Volta'' and CNISM,
  via Bassi 6, I-27100 Pavia, Italy}

\author{R. Demkowicz-Dobrza\'nski}

\affiliation{Center for Theoretical Physics of the Polish Academy of
  Sciences, Aleja Lotnik\'ow 32/46, 02-668 Warszawa, Poland.} 

\author{P. Perinotti} 

\affiliation{QUIT Group, Dipartimento di Fisica ``A. Volta'' and CNISM,
  via Bassi 6, I-27100 Pavia, Italy}

\author{M. F. Sacchi}

\affiliation{QUIT Group, Dipartimento di Fisica ``A. Volta'' and CNISM,
  via Bassi 6, I-27100 Pavia, Italy}

\affiliation{CNR - Istituto Nazionale per la Fisica della Materia, 
Unit\`a di Pavia, Italy.}  

\begin{abstract}
We address the problem of removing correlation from sets of states 
while preserving as much local quantum information as possible. 
We prove that states obtained from universal cloning can
only be decorrelated at the expense of complete erasure of 
local information (i.e. information about the copied state).
We solve analytically the problem of decorrelation for two qubits and two qumodes  
(harmonic oscillators in Gaussian states), and provide sets 
of decorrelable states and the minimum amount of noise to be added 
for decorrelation.
\end{abstract}

\maketitle 

The laws of quantum mechanics impose a number of restrictions on the
processing of \emph{quantum information}.  Examples of such impossible
tasks are provided by the famous no-cloning theorem \cite{Wootters82}
or by the theorem on non-existence of the universal-NOT gate
\cite{Buzek99}.  Despite their discouraging appearance, such
limitations can sometimes be proved useful. This is the case with the
no-cloning theorem which is at the core of quantum cryptography,
as it prevents an eavesdropper from creating perfect copies of a transmitted 
quantum state.

In this Letter we attempt to broaden our understanding of the
limitations imposed on the quantum information processing, by
investigating the possibility of \emph{decorrelating quantum states}
-- i.e. removing unwanted correlations between quantum subsystems
while preserving local information encoded in each of them.

To be more specific, we say that an operation $\mathcal{D}$
\emph{faithfully decorrelates} an $N$-partite state $\rho$ if the
following equation holds:
\begin{equation}
\label{eq:decorgeneral} \mathcal{D}\left(\rho \right) =
\rho_1 \otimes \dots \otimes \rho_N, 
\end{equation} 
where $\rho_i$ is the $i$-th party reduced density matrix of $\rho$.
Now, the problem of decorrelability can be stated as follows: given a
set of states $S = \left\{ \rho \right\}$, we ask whether there exists
a \emph{single} physically realizable operation (completely positive
map) $\mathcal{D}$ that satisfies (\ref{eq:decorgeneral}) for every
state $\rho$ in $S$.

Analogously as in the case of the no-cloning theorem, the answer will
strongly depend on the chosen set of states. In particular, if the set
$S$ consists of only one element $\rho$, then the problem of
decorrelability is trivial.  One can always choose $\mathcal{D}$ to be
a map producing $\otimes _{i=1}^N \rho_{i}$ out of any input.  Such a
map is completely positive and hence every single-element set is
decorrelable.

Moving to the other extreme, and asking whether a set $S$
consisting of all density matrices is decorrelable, one finds out that
due to linearity of quantum mechanics it is not
\cite{Terno}. Actually, from the proof of \cite{Terno} one can easily
draw a stronger conclusion, namely:

If a set $S$ contains the states $\rho'$, $\rho''$ and their convex
combination $\lambda \rho' +(1-\lambda) \rho''$, and the reduced
states of $\rho'$ and $\rho''$ are different at least for two parties,
then faithful decorrelability of the set $S$ is impossible.
 
Moreover, in \cite{Terno} nondecorrelability of certain two-element
sets was shown using the fact that after decorrelation
distinguishability of states cannot increase (see also \cite{Mor} for
some results on disentangling rather than decorrelating states).
Apart from the above particular cases, very little is known on the
decorrelability of general sets of quantum states.  In this Letter we
search for explicit solutions to the decorrelation problem in
interesting settings.

The key factor for deciding on decorrelability and nondecorrelability
is the choice of the set of states. In this Letter such a choice is 
motivated by the need of considering the problem of  decorrelation 
in information-processing context.  We stress that we decorrelate quantum
\emph{states} by keeping the quantum \emph{signals}. We propose the
following paradigm.

Consider an $N$-partite \emph{correlated} ``seed'' state $\rho$, which
should be regarded as the initial state where information is encoded.
Let $U_g$ be a unitary representation of a group $G$, acting on the
Hilbert space of a single party. The representation describes the
encoding procedure of a piece of information (the group element $g$)
on the state of a subsystem.  Acting with unitary operations $U_{g_1}
\otimes \dots \otimes U_{g_N}$ on the seed state $\rho$ should be
regarded as encoding $N$ pieces of information ($N$ signals
\cite{nota1}) ${\boldsymbol g}=(g_1, \dots, g_N)\in G^N$:
\begin{equation}
\rho_{\boldsymbol g}\ceq U_{g_1}\otimes \dots 
\otimes U_{g_N} \rho U_{g_1}^\dagger \otimes \dots \otimes U_{g_N}^\dagger.
\end{equation}
The above state is clearly correlated due to the correlation of the
seed state $\rho$.  The problem of decorrelation is to find a
single operation $\mathcal{D}$ that would decorrelate [see Eq.
(\ref{eq:decorgeneral})] all states belonging to the set:
\begin{equation}
\label{eq:ensemble}
S(U_g,\rho)=\{\rho_{\boldsymbol g},
\forall {\boldsymbol g}\in G^N\}.  
\end{equation} 
If we have additional constraints on the signals (e.g. we know that
they are identical $g_1=\dots=g_N$) the above set is smaller, and
eventually the problem of decorrelation becomes easier. Notice that
the reduced density matrices of $\rho_{\boldsymbol g}$ are related to
the reduced density matrices of $\rho$ by $U_{g_i}\rho_i
U_{g_i}^\dagger$, and as a result the decorrelated state still carries
the same signals as the initial one.  We stress that we do not deal
here with decorrelation of \emph{signals}, but rather with
decorrelation of states carrying them (hence, there is no
contradiction in performing decorrelation and still claiming, e.g.,
that the encoded signals are identical).

To motivate our work further let us recall some facts about cloning
and state estimation.  We know that quantum information cannot be
copied or broadcast exactly, due to the no-cloning theorem.
Nevertheless, one can find approximate optimal cloning operations
which increase the number of copies of a state at the expense of the
quality.  In the presence of noise, however, (i. e. when transmitting
``mixed'' states), it can happen that we are able to increase the
number of copies without losing the quality, if we start with
sufficiently many identical originals. Indeed, it is even possible to
{\em purify} in such a broadcasting process---the so-called {\em
  super-broadcasting} \cite{our}.  Clearly, a larger number of copies
cannot increase the available information about the original input
state, and this is due to the fact that the final copies are not
statistically independent, and the correlations between them limit the
extractable information \cite{estcor}.  It is now natural to ask if we
can remove such correlations and make them independent again (notice
that in this decorrelation problem, the signals $g_i$ -- which in this
case correspond to the cloned states -- are identical).  Clearly, such
quantum decorrelation cannot be achieved exactly, otherwise we would
increase the information on the state. A priori it is not excluded,
however, that it is possible to decorrelate clones at the expense of
introducing some additional noise. One of the results of this Letter
is that clones by universal cloning cannot be decorrelated even within
this relaxed condition.  Apart from this negative result, we provide
in this Letter examples of states for which decorrelation is possible.

Thanks to the structure of the set of states (\ref{eq:ensemble}) that we want to decorrelate, 
a covariant decorrelation must satisfy the following conditions: 
(i) $\mathcal{D}$ decorrelates the seed state; (ii) $\mathcal{D}$  
fulfills the covariance condition:
\begin{equation}
\begin{split}
  \mathcal{D}\left(U_{g_1}\otimes \dots \otimes U_{g_N}
    \rho U_{g_1}^\dagger \otimes \dots \otimes U_{g_N}^\dagger \right)=\\
  U_{g_1}\otimes \dots \otimes U_{g_N} \mathcal{D}\left(\rho \right)
  U_{g_1}^\dagger \otimes U_{g_N}^\dagger.
\end{split}
\end{equation}
We will more generally consider decorrelation with additional noise on
the output local states, namely
\begin{equation}
\label{eq:decorseed}
\mathcal{D}(\rho)=\tilde{\rho}_1 \otimes \dots \tilde{\rho}_N,
\end{equation}
where $\tilde{\rho}_i \neq \rho _i$.  As a result, subsystems are
still perfectly decorrelated, but some information about reduced
density matrices is lost.  Additionally, in what follows we will
assume that the seed state is permutationally invariant---in other
words we treat all encoded signals on equal footing. This simplifies
the situation since in this case all single party reduced density
matrices of the seed state are equal and the same should hold for the
noisy reduced density matrices after decorrelation.  This assumption
allows us, without loss of generality, to impose permutational
covariance on the decorrelating operation $\mathcal{D}$.

We now present the solution for some interesting bipartite situations.
We analyze \emph{qubits}, in which information is encoded through
general unitaries in $\SU 2$, and \emph{qumodes} (harmonic oscillators
in Gaussian states), with information encoded by the representation of
the Weyl-Heisenberg group of displacements. In our analysis we
consider the two situations in which the unitaries representing
signals on the two systems are either independent or equal. The latter
case is relevant for the decorrelability of output states of cloning
and broadcasting machines. It turns out that decorrelation is indeed
possible in some cases, at the expense of increasing local noise. The
optimal decorrelating map adding the minimum amount of noise is
derived in the qubit case, and the optimal depolarization factor is
evaluated as a function of the input seed state. For Gaussian states
we show that it is always possible to erase correlations by means of a
suitable Gaussian map.

\par Consider a couple of qubits $A$ and $B$. Permutational invariance
of the seed state means that it is block diagonal with respect to
singlet and triplet subspaces.  For qubits the state is conveniently
described in the Bloch form.  Without loss of generality we may assume
that the reduced density matrices $\rho_A=\rho_B=\frac12(\openone +
\eta \sigma_z)$ of the seed state $\rho_{AB}$ are diagonal in the
$\sigma_z$ basis.  The information $(\alpha,\beta)$ is encoded via the
action of $U(\alpha) \otimes U(\beta)$:
\begin{equation}
\rho_{AB}(\alpha,\beta)=U(\alpha)\otimes U(\beta) \rho_{AB}
U(\alpha)^\dagger \otimes U(\beta) ^\dagger \;, 
\end{equation}
where $\alpha $ and $\beta $ are elements of $\SU 2$.  
In other words it is encoded on the direction of the
Bloch vectors $\vec{n}_A(\alpha)$ and $\vec{n}_B(\beta)$ of the
marginal states
\begin{equation}
\begin{split}
  \rho_A(\alpha)=&\Tr_B[\rho_{AB}(\alpha,\beta)]=\tfrac{1}{2}(\openone
  + \eta\vec{n}_A(\alpha)\cdot\vec{\sigma}),
  \\
  \rho_B(\beta)=&\Tr_A[\rho_{AB}(\alpha,\beta)]=\tfrac{1}{2}(\openone
  + \eta \vec{n}_B(\beta)\cdot\vec{\sigma}),
\end{split}
\end{equation}
where $\vec\sigma=(\sigma_x,\sigma_y,\sigma_z)$ is the vector of Pauli
matrices.  Covariance of the decorrelation map means that the
direction of the Bloch vectors $\vec{n}_A(\alpha)$ and
$\vec{n}_B(\beta)$ should be preserved in the output states, i.~e.
\begin{equation}
\begin{split}
  & \tilde{\rho}_A(\alpha)=\frac{1}{2}(\openone + \tilde \eta
  \vec{n}_A(\alpha)\cdot\vec{\sigma}),\\
  & \tilde{\rho}_B(\beta)=\frac{1}{2}(\openone +\tilde \eta
  \vec{n}_B(\beta)\cdot\vec{\sigma}),
\end{split}
\end{equation}
namely only the length of the Bloch vector (i.~e. the purity of the
state) is changed $\eta\to\tilde\eta$. The additional noise of the
output states corresponds to a reduced length of the Bloch vector
$\tilde \eta <\eta$. The directions of the Bloch vectors $\vec
n_A(\alpha)$ and $\vec n_B(\beta)$ are completely arbitrary. The
optimal decorrelation map will maximize the length $\tilde \eta$ of
the Bloch vector, namely it will produce the highest purity of
decorrelated states. It can be shown \cite{unp} that the general form
of a two-qubit channel $\map{D}$ covariant under $U(\alpha) \otimes
U(\beta)$ and invariant under permutations of the two qubits can be
parameterized with three positive parameters only (effectively two due
to normalization)
\begin{equation}
\map{D}(\rho_{AB})= a \rho_{AB} + b \map{D}_1(\rho_{AB})+c
\map{D}_2(\rho_{AB}),
\end{equation}
where $\map{D}_1$ and $\map{D}_2$ are given by
\begin{align}
  \map{D}_1(\rho_{AB})=&\tfrac{1}{3}\left(
    \rho_A \otimes \openone + \openone \otimes \rho_B - \rho_{AB} \right) \;,\\
  \map{D}_2(\rho_{AB})=&\tfrac{1}{9}\left(4 \openone \otimes \openone
    -2 \rho_A \otimes \openone- 2 \openone \otimes \rho_B +
    \rho_{AB}\right) \;,
\end{align}
and the trace preserving condition gives $a+b+c=1$. This is a very
restricted set of operations, due to the fact that the
covariance condition is very strong. 
 As a consequence the condition for decorrelating the seed state
\begin{equation}
\label{eq:condseed}
\mathcal{D}(\rho_{AB})=\tilde{\rho}^{\otimes 2}=\left[\frac12(\openone + \tilde \eta \sigma_z)\right]^{\otimes 2}
\end{equation}
cannot be satisfied for a generic 
seed state $\rho_{AB}$ (apart
from the trivial decorrelation to a maximally mixed state).

\begin{figure}[ht]
\includegraphics[width= 0.5\textwidth]{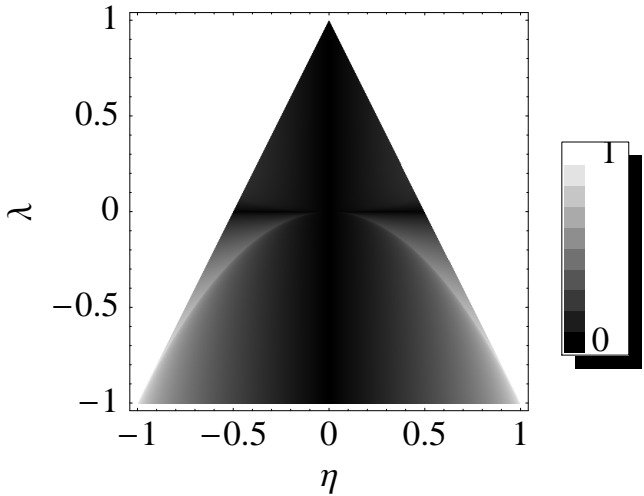}
\caption{Length $\tilde \eta $ of the Bloch vectors of the decorrelated
  states of two qubits starting from the joint state in Eq.
  (\ref{decorrelable}). The plot depicts the maximal achievable
  $\tilde \eta $ in gray scale versus the parameters $\eta$ and $\lambda$ of the
  input state.}
\label{f:eta}
\end{figure}

\par\noindent  The seed states for which nontrivial decorrelation is possible 
[which means that we can find such $a,b,c$ and $\tilde \eta >0$
satisfying Eq. (\ref{eq:condseed})]  
have the form  \cite{unp}
\begin{equation}\label{decorrelable}
  \rho_{AB}=\tfrac{1}{4}\left[\openone \otimes \openone + \eta
    (\sigma_z \otimes \openone + \openone \otimes \sigma_z) - \lambda
    \sigma_z \otimes \sigma_z \right].
\end{equation}
We emphasize that for a generic seed state $\rho_{AB}$ one can reduce
correlations, but only sets arising from the seed state of the form
(\ref{decorrelable}) can be decorrelated completely in a nontrivial
way (apart from the cases when $\eta=0$ or $\lambda=0$).  The noise of
the decorrelated states depends on parameters $\eta$ and $\lambda$ as
depicted in Figure \ref{f:eta}.

\par In order to study the decorrelability of the output states of
cloning machines, we consider now the case where the same unitary is
encoded on the two qubits (identical signals). Differently from the
case of independent signals, $\mathcal{D}$ has to be covariant with
respect to $U(\alpha)^{\otimes 2}$, which is a weaker condition than
covariance with respect to $U(\alpha) \otimes U(\beta)$. Using the
methods from \cite{our} one can parameterize these class of operations
with six parameters $s_{j,l,J}$ satisfying two constraints, so
effectively one enjoys a four parameter freedom on covariant
operations. Thanks to this larger freedom it can be shown \cite{unp}
that the decorrelation condition
$\mathcal{D}(\rho_{AB})=\tilde{\rho}^{\otimes 2}$ is non trivially
satisfied (i.e. for $\tilde \eta >0$) for a generic state $\rho_{AB}$
which is diagonal in the singlet triplet basis. Such a state can be
written in the form:
\begin{equation}
\rho_{AB}=p\ket{\Psi^-}\bra{\Psi^-}+(1-p)\rho_{\textrm{sym}},
\end{equation}
where $\rho_{\textrm{sym}}$ is a state supported on the triplet
(symmetric) subspace, and $\ket{\Psi^-}$ is the singlet state.
Analogously to Eq. (\ref{decorrelable}), $\rho_{\textrm{sym}}$ can be
written with the help of Pauli matrices:
\begin{equation}
\label{eq:sym}
\begin{split}
\rho_{\textrm{sym}}=\frac{1}{4}[\openone\otimes \openone + 
\eta (\sigma_z \otimes \openone + \openone \otimes \sigma_z) + \\
(1+ \lambda)/2\ (\sigma_x \otimes \sigma_x + \sigma_y \otimes \sigma_y ) 
- \lambda \sigma_z \otimes \sigma_z]
\end{split}
\end{equation}


\par \noindent The only states that cannot be nontrivially
decorrelated are those with either $p=1$ or $\eta=0$ or
$\lambda=-1/3$. For $p=0$ (i.e. for seed
states supported on the symmetric space) the plot for achievable $\tilde{\eta}$ is analogous to Fig. 1, 
but now the black horizontal line containing non-decorrelable states lays at $\lambda=-1/3$.
Interestingly, these non-decorrelable states are states which can be obtained via universal
cloning machines producing two copies out of one copy of a qubit
state. Hence, clones obtained by $1$-to-$2$ universal cloning cannot be
nontrivially decorrelated. 

Such a result can be shown in general for
$N$-to-$M$ universal cloning of $d-$dimensional systems (\emph{qudits}) along the following lines.  W.l.o.g. we
can restrict ourselves to $M=N+1$ and consider cloning of pure states \cite{nota2}.
The universal covariance of the cloning and decorrelation procedure implies that for every pure state $\ket{\phi}$ of a qudit the desired transformation has the form:
\begin{equation}
  \label{eq:phaseoutput} 
\Lambda[(\ket{\phi}  \bra{\phi})^{\otimes N}] = \left(\eta \ket{\phi} \bra{\phi}+\frac{1-\eta}{2}\openone \right)^{\otimes N+1}
\end{equation}
The transformation (\ref{eq:phaseoutput}) is only possible for
$\eta=0$. Indeed, let us consider these transformation for 
``equatorial states'' i.e. 
$\ket{\phi}=(\ket{0} + e^{i\phi} \ket{1})/\sqrt{2}$ where $\ket{0}$, $\ket{1}$ are some arbitrary orthogonal states. If $\eta \neq 0$ entries of the operator on the right hand side of (\ref{eq:phaseoutput}) are polynomials of degree
at most $N+1$ of $e^{\pm i \phi}$ (and some entries actually achieve this degree ). On the other hand thanks to linearity of $\Lambda$ 
the entries on left hand side are polynomials of degree at most $N$ of $e^{\pm i \phi}$. Since equality (\ref{eq:phaseoutput}) should be satisfied for all
phases $\phi$ we arrive at a contradiction, since no polynomial of degree 
$N$ can be equal to a polynomial of degree $N+1$ in an infinite number of points.
 The above reasoning holds
true also for asymmetric cloning with different $\eta$ for each
output, where one can prove that at least one coefficient $\eta$
must be null \cite{nota3}.

We consider now the case of decorrelation for qumodes. For a
couple of qumodes in a joint seed state $\rho_{AB}$ the information
$(\alpha,\beta)$ (with $\alpha$ and $\beta$ complex) is encoded as
follows
\begin{equation}\label{qstate}
  D(\alpha )\otimes D(\beta)\rho_{AB} D(\alpha )^\dagger \otimes D(\beta)^\dagger,
\end{equation}
$D(z)=\exp(za^\dag-z^*a)$ for $z\in{\mathbb C}$ denoting a single-mode
displacement operator, $a$ and $a^\dag$ being the annihilation and
creation operators of the mode.  Here we show that it is always possible to decorrelate any joint
state of the form (\ref{qstate}), with $\rho _{AB}$ representing a
two-mode Gaussian state, namely
\begin{eqnarray}
\rho_{AB} = \frac {1}{\pi ^4}\int d^4\vec{q} \,e^{-\frac 12 \vec{q}^T\vec{M}\vec{q}}D(\vec{q})\;,
\end{eqnarray}
where $\vec{q}=(q_1,q_2,q_3,q_4)$, $D(\vec{q})=D(q_1 +i q_2)\otimes D(q_3 +i q_4)$, 
and $\vec{M}$ is the $4 \times 4$ (real, symmetric, and positive)
correlation matrix of the state, that satisfies the Heisenberg uncertainty relation  \cite{simon} 
$\vec{M} + \frac i 4 \vec{\Omega }\geq 0$, 
with $\vec {\Omega }=\oplus _{k=1}^2 \vec{\omega }$ and $\vec{\omega }=
\left (\begin{array}{cc}
0 & 1\\ -1 &0 
\end{array}
\right )$. 

A Gaussian
decorrelation channel covariant under $D(\alpha)\otimes D(\beta)$ is
given by
\begin{equation}\label{qdecorr}
  \map{D}(\rho)= \frac {\sqrt{\hbox{det}\vec{G}}}{(2\pi)^2}\int d^4
  \vec{x}\,e^{-\frac 12\vec{x}^T\vec{G}\vec{x}} D(\vec{x}) \rho D^\dag (\vec{x}),
\end{equation}
with suitable positive matrix $\vec G$ \cite{unp}, and the resulting state $\map
{D}(\rho _{AB})$ is still Gaussian, with a new block-diagonal
covariance matrix $\widetilde{\vec M}$, thus corresponding to a
decorrelated state.

A special example of Gaussian state of two qumodes is the so-called
{\em twin beam}, which can be generated in a quantum optical lab by
parametric down-conversion of vacuum. In this case $\vec M$ is given
by
\begin{eqnarray}
\vec M= \frac {1+ \lambda ^2}{1- \lambda ^2} \openone - \frac {2 \lambda }{1-
  \lambda ^2} \left ( 
\begin{array}{cc}
0 & \sigma _z \\ \sigma _z &0 
\end{array}
\right )\;,
\end{eqnarray}
with $0\leq \lambda <1$. 
The map (\ref{qdecorr}) with 
\begin{eqnarray}
\vec G= \frac {2 \lambda }{1- \lambda ^2} 
\left [\openone + \left ( 
\begin{array}{cc}
\varepsilon  & \sigma _z \\ \sigma _z & \varepsilon 
\end{array}
\right )\right ]\;,
\end{eqnarray}
and arbitrary $\varepsilon >0$, provides two decorrelated states with
$\widetilde{\vec M} = (\frac{1+ \lambda }{1- \lambda }+\varepsilon )
\openone $, which correspond to two thermal states with mean photon
number $\bar n= \frac{\lambda }{1-\lambda }+\frac {\varepsilon}{2}$
each. Since the channel in Eq.~\eqref{qdecorr} is covariant also for
$D(\alpha)^{\otimes 2}$, the above derivation then holds for the case
of encryption with the same unitary on both qumodes as well.

\par The striking difference between the qubit and the qumode cases is
that for qubits only few states can be decorrelated, whereas for
qumodes any joint Gaussian state can be decorrelated. This is due to
the fact that the covariance group for qubits comprises all local
unitary transformations, whereas for qumodes includes only local
displacements, which is a very small subset of all possible local
unitary transformations in infinite dimension. In particular it can be
checked that unlike qudits, states obtained via Gaussian cloning of
coherent states can be decorrelated and the no-go proof valid for
finite dimensional cases does not apply here.

\acknowledgments This work has been supported by Ministero Italiano
dell'Universit\`a e della Ricerca (MIUR) through PRIN 2005 and the
Polish Ministry of Scientific Research and Information Technology
under the (solicited) grant No.  PBZ-Min-008/P03/03.


\begin{thebibliography}{99}
\bibitem{nota1} The word signal suggests sometimes a sequence of
  pieces of information being transmitted. In our case this will
  simply correspond to repeated preparation of the state
  $\rho_{\boldsymbol g}$ with varying set ${\boldsymbol
    g}=(g_1,\dots,g_N)$ in each shot.
\bibitem{nota2} For the proof of impossibility it is enough to
  consider $M=N+1$ since if cloning without correlations were possible
  for $M'>N+1$, then by partial trace over $M'-N-1$ output copies we
  would obtain a cloning operation with $M=N+1$. Moreover, we can
  restrict attention to pure states, since if cloning were possible
  for mixed states, then it would be possible for pure states, too.
  Indeed one could apply depolarizing channels to the input pure
  states before cloning.
\bibitem{nota3} Strictly speaking  the above statement can be proved without
the use of the covariance condition (see \cite{unp}). Covariance condition was 
used here for the sake of simplicity and consistency with the general approach taken in the Letter.
\end{thebibliography}
\end{document}